\documentclass[a4paper,11pt]{article}
\usepackage{pos}
\usepackage{graphicx}
\usepackage{subcaption}

\title{The simulation chain for the Terzina Cherenkov telescope on board the NUSES space mission}
 \ShortTitle{The simulation chain for the Terzina Cherenkov telescope on board NUSES}

\author*[a,b]{C. Trimarelli}

\affiliation[b]{GSSI - Gran Sasso Science Institute, Via F. Crispi 7, I-67100, L'Aquila, Italy}
\affiliation[c]{INFN - Laboratori Nazionali del Gran Sasso, Via G. Acitelli 22, I-67100, Assergi, L'Aquila, Italy}

\emailAdd{caterina.trimarelli@gssi.it}

\abstract{The Terzina telescope is designed to detect ultra-high energy cosmic rays (UHECRs) and Earth-skimming neutrinos from a 550~km low-Earth orbit (LEO) by observing Cherenkov light emitted by Extensive Air Showers (EAS) in the Earth’s atmosphere pointing towards the telescope and in the field of view. In this contribution, a simulation chain for the Terzina telescope on board the NUSES mission will be presented. The chain encompasses all stages of the detection process, from event generation and EAS modelling with CORSIKA and EASCherSim to Geant4-based simulations of the telescope’s geometry and optics, followed by modelling of the trigger system and silicon photomultiplier (SiPM) response. The Geant4 module includes the real CAD model of the telescope structure and optical components, with aspherical lenses manually implemented to ensure accurate representation of the optical efficiency and point spread function in Geant4. The expected background for this experiment was evaluated in detail using the data from the Defense Meteorological Satellite Program (DMSP) satellite for city lights contributions, the Lunar Irradiance Model of the European Space Agency (LIME) for the Moonlight, the effect of auroral emissions, as well as radiation sources responsible for dose deposition in the telescope components. Radiation tolerance was assessed by simulating the dose from charged particles using SPENVIS particle flux predictions. SiPM characterisation was used to simulate the SiPM response within the simulation, as well as the realistic signal shape from the preamplifier and conditioning electronics chain. This comprehensive pipeline, developed using modular C++ code and Python tools for event analysis and reconstruction, produces detailed performance assessments of a telescope operating in a LEO mission but can be adapted for any high altitude Cherenkov telescope, making it a versatile tool for future observatory designs. The possibility of modelling balloons in the atmosphere has also been developed.}

\ConferenceLogo{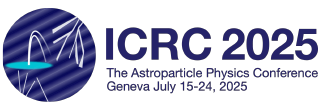}

\FullConference{39th International Cosmic Ray Conference (ICRC2025)\\
 15–24 July 2025\\
Geneva, Switzerland\\}

\begin{document}
\maketitle

\section{Introduction}


The NUSES mission~\cite{Trimarelli:2024rdb} aims to open new directions for space-based astroparticle physics with a satellite in a quasi-polar low-Earth orbit (LEO) at 550~km. Among its payloads~\cite{Burmistrov_2023, DeMitri:2023ejv}, the Terzina telescope is dedicated to the first observation from space of Cherenkov light produced by extensive air showers (EAS) developing along the Earth’s limb. Depending on the direction and origin of the incoming primary particle, Terzina can observe two different types of events by looking just above or just below the Earth’s limb. It detects Cherenkov-emitting extensive air showers from ultra-high energy cosmic rays when pointing above the limb, and upward-moving showers from Earth-skimming neutrinos when pointing below the limb. Terzina is tilted at 67.05$^{\circ}$ relative to the nadir to maintain continuous pointing towards the Earth’s night-side limb. The telescope employs a compact Schmidt-Cassegrain optical design, combining mirrors and a correcting lens to minimise aberrations while focusing light onto a flat focal plane array, with an effective collecting area of 0.09~m$^2$, a focal length of 922~mm, and a total field of view (FoV) of $7.4^\circ \times 3.0^\circ$. As a pathfinder, Terzina demonstrates the feasibility of observing Cherenkov light from space and provides essential input for future large-scale missions, such as POEMMA\cite{Olinto,POEMMA:2020ykm}.

To support this goal, a dedicated simulation chain has been developed. 
It comprises the generation and propagation of the EAS signal using CORSIKA~\cite{Heck:1998vt} 
and EASCherSim~\cite{C1,C2}\footnote{EasCherSim: \url{https://pypi.org/project/easchersim/}} 
(see Section~\ref{Cherenkovsignal}); the detailed detector and optical modelling, 
based on the real CAD geometry implemented in Geant4~\cite{GEANT4} 
(Section~\ref{sec:terzinag4}); and the simulation of the trigger system and SiPM response, 
including realistic electronic signals (Section~\ref{terzinareadoutsim}).  The expected background and radiation environments have been evaluated using external datasets such as Defense Meteorological Satellite Program (DMSP)~\cite{Lieske1981DMSPPS} for city lights, Lunar Irradiance Model of the European Space Agency (LIME)~\cite{LIME1} for moonlight, and SPENVIS~\cite{SPENVIS} for radiation flux predictions, while detector characterisation was included to reproduce realistic noise and degradation effects (see \cite{JCAP_p,shideh}). This modular pipeline enables comprehensive performance studies of the Terzina telescope in LEO and provides a flexible framework that can be adapted to other Cherenkov instruments. In the following, we present its main components and illustrate how each stage contributes to the overall simulation and analysis workflow.

\section{Terzina Simulation Chain}

\begin{figure}[ht!]
    \centering      
    \includegraphics[width=0.95\textwidth]{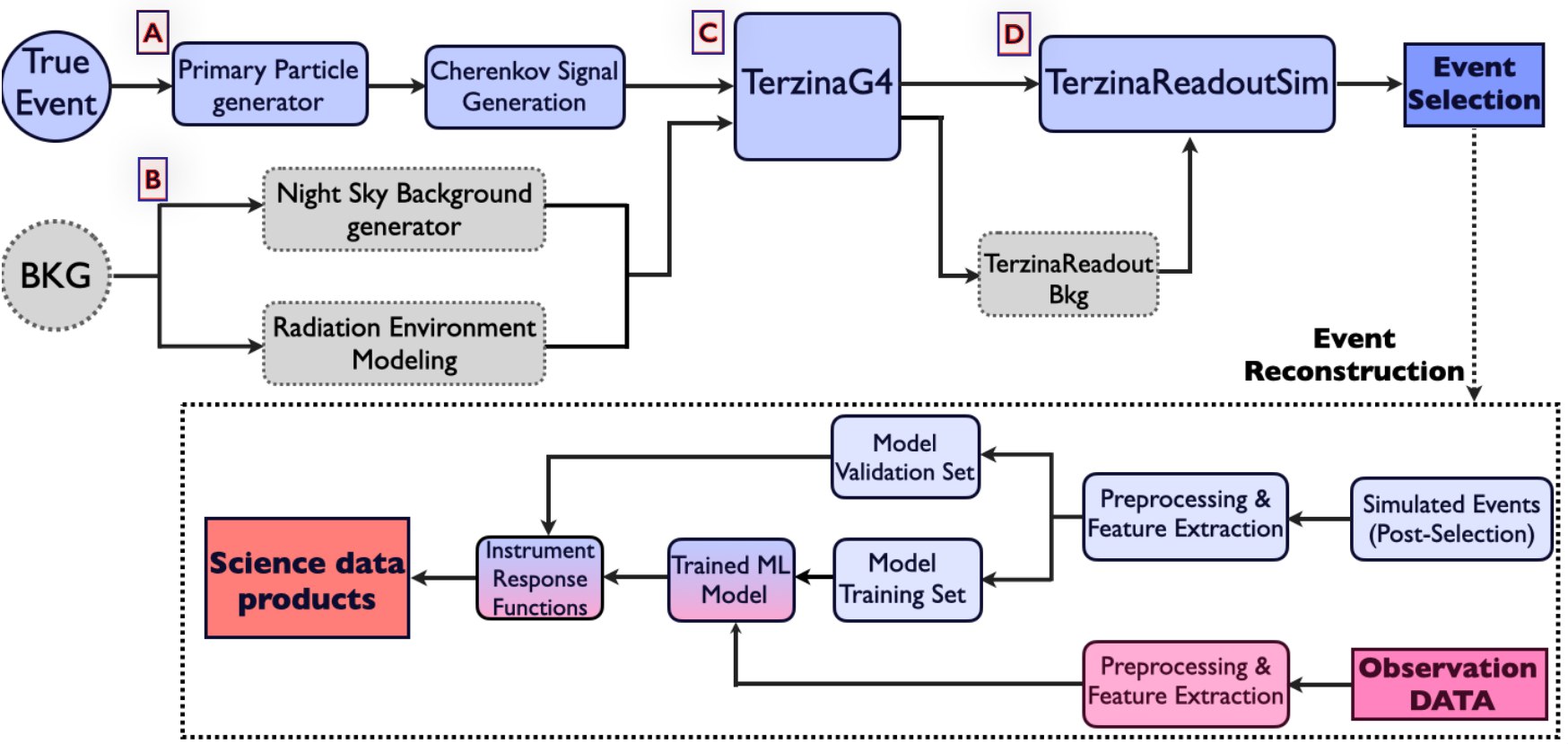}
    \caption{Schematic overview of the simulation and analysis workflow. The upper part illustrates the generation of true events and background (BKG) contributions, their propagation through detector simulation (TerzinaG4 and TerzinaReadoutSim), and the event selection process. The lower part represents the event reconstruction workflow, including preprocessing and feature extraction from both simulated and observed data, model training and validation for the machine learning model, and the production of science data products using instrument response functions.}
    \label{fig:simchain}
\end{figure}

The telescope’s simulation chain, a schematic overview is shown in Figure~\ref{fig:simchain}, is structured into multiple stages that reproduce both the Cherenkov signatures and the various background contributions that can affect their detection. It is used to define event selection criteria and to estimate backgrounds for the LEO mission, such as the characterisation of the night glow background (NGB) and the assessment of the radiation hardness of detector components, thus providing a reliable assessment of the detector performance.

The full simulation of the Terzina telescope can be  divided into two main stages:
\begin{enumerate}
\item  photon and particle generation: this stage simulates the observable features for Terzina, namely Cherenkov photons as measurable signal (letter A in the scheme) and background sources (letter B in the scheme) and their impact on the performance. The signal generation includes the simulation of the geometrical filtering of the cosmic ray and neutrino tracks through the \textsc{Cosmic Proton Generator (CPG)} and the \textsc{Neutrino Generator (nuG)} and the emission and the propagation of the Cherenkov signal using both CORSIKA and EASCherSim.

The background generation component (letter B in the scheme) models the background contributions, including night sky illumination from city lights and moonlight, as well as radiation effects on the detector. External datasets such as DMSP, LIME, and SPENVIS provide input to estimate photon rates, dose deposition, and SiPM degradation over time. Dedicated Python tools have been developed to process these datasets and compute realistic photon background rates, taking Terzina’s orbit into account and pixel field of view. They provide per-pixel photon fluxes with full angular and temporal distributions, including timestamps, FoV pointing coordinates, photon counts, and spectral information. The outcomes and more details of these tools are described in \cite{TerzinaICRC2025,JCAP_p,shideh}.

\item Detector response simulation: this stage models how the telescope records and processes incoming photons. It includes the \textsc{TerzinaG4} (Geant4-based) module (letter C in the diagram) which simulates the interaction of all incoming photons with the telescope, including Cherenkov signal photons and background photons, and is also used to evaluate radiation dose in the detector. The \textsc{TerzinaReadoutSim} module (letter D in the diagram) simulates the readout chain, including waveform generation, SiPM response with noise and gain variations, front-end electronics, and trigger logic. A configuration file allows setting parameters such as mission duration, photon rates, radiation-dependent dark count evolution, preamplifier and waveform settings, and trigger thresholds.

Each of these components will be discussed in detail in dedicated sections below. The part of the workflow that is still under development, covers event reconstruction—including preprocessing, feature extraction, and model training—but will not be described here.

\end{enumerate}

\begin{figure}[ht]
    \centering

    \begin{subfigure}[b]{0.45\textwidth}
        \centering
        \includegraphics[width=\textwidth]{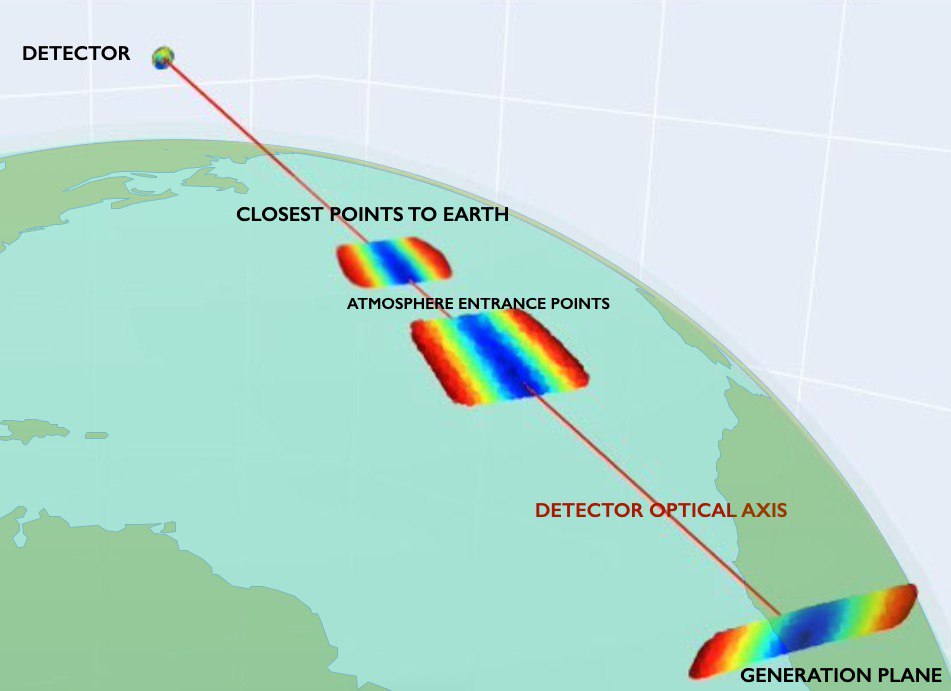}
        \caption{}
        \label{fig:CPG}
    \end{subfigure}
    \hfill
    \begin{subfigure}[b]{0.43\textwidth}
        \centering
        \includegraphics[width=\textwidth]{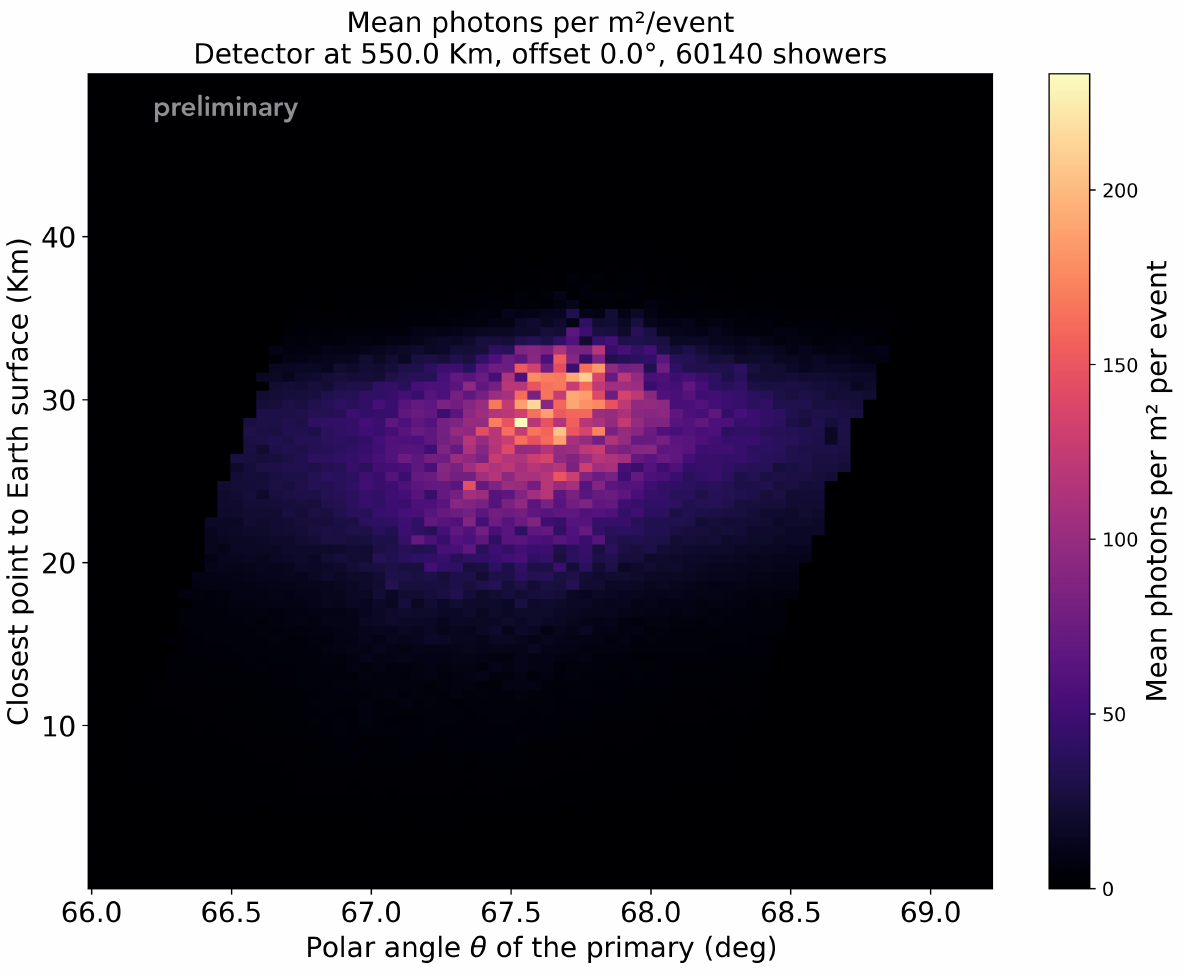}
        \caption{}
        \label{fig:SecondImage}
    \end{subfigure}

    \caption{Left: Geometry of the tracks generator. In the 3D-sphere configuration, the generation plane, the closest position of the primaries to Earth, the detector position are shown. Right: Distribution of Cherenkov photons produced by a 100 PeV proton-induced EAS as a function of the primary particle polar angle and the closest position of the protons to Earth.}
    \label{fig:CombinedFigure}
\end{figure}

\section{Cherenkov signal at LEO}
\label{Cherenkovsignal}
The first step in the simulation chain is the pre-production phase, where the tracks of the primary particles that induce EAS are generated using CPG and nuG. This stage performs a geometrical pre-selection, retaining only candidate tracks compatible with Terzina’s geometrical aperture. The two generators are based on the same concepts but considering, in the neutrino case, the Earth emergence probability of the tau produced by the neutrino interaction with the Earth’s crust. Since the nuG is under development the results shown in this work rely on the CPG. 
The cosmic rays originate from a generation plane that is positioned perpendicular to the detector’s optical axis. Once generated, the particles travel towards the telescope, but before reaching it, they interact with the atmosphere, producing EAS. The point of emission, which represents the closest approach of the cosmic rays to the Earth, must be within a range of 0 to 100~km. The impact parameter, defined as the distance, on the detection plane, between the telescope optical axis and the intersection point of the track, is constrained between 0 and 50~km. To ensure accurate simulation in subsequent steps, another crucial constraint is that the angle between Terzina and the cosmic ray tracks must be between 180$^{\circ}$ and 180$^{\circ}$-FoV$_x/2$. Figure~\ref{fig:CPG} illustrates the geometry employed in this stage of the simulation. 
After applying all selection criteria, the key characteristics of the remaining primary cosmic ray tracks, including their angles and altitudes, are recorded and used as input for the next stages of the simulation, which involve EASCherSim and Geant4. 
The optical properties of the Cherenkov signal generated by proton interactions with the atmosphere are characterised using EASCherSim, a dedicated optical simulation tool that models the production and propagation of Cherenkov light to the telescope.
As an input, for each selected primary track one should provide the viewing angle that is strictly connected with the polar angle of the track, the detector altitude that changes accordingly to the impact parameter, the primary energy and the charged particle distribution of the corresponding shower. The charged particle distributions are taken from simulations of proton extensive air showers produced with CORSIKA\footnote{Corsikav77500 with Fluka2021-2 EPOS-LHC (CURVED THIN SLANT)}. Fluctuations of the first interaction point have been accounted for by using different showers with the same primary energy.
Detailed calculations of photon density, spectral composition, angular distribution, and temporal evolution, which are crucial for predicting the signal observed by the Terzina telescope are provided. In Figure~\ref{fig:SecondImage}, the result of the complete first step of the simulation chain is shown for 60,000 selected Terzina proton events with energies of 100~PeV, illustrating the photon density as a function of atmospheric height and the polar angle of the track.


\section{TerzinaG4: optics and detector Simulation}
\label{sec:terzinag4}
\begin{figure}[t]
 \centering
  \includegraphics[width=1.\textwidth]{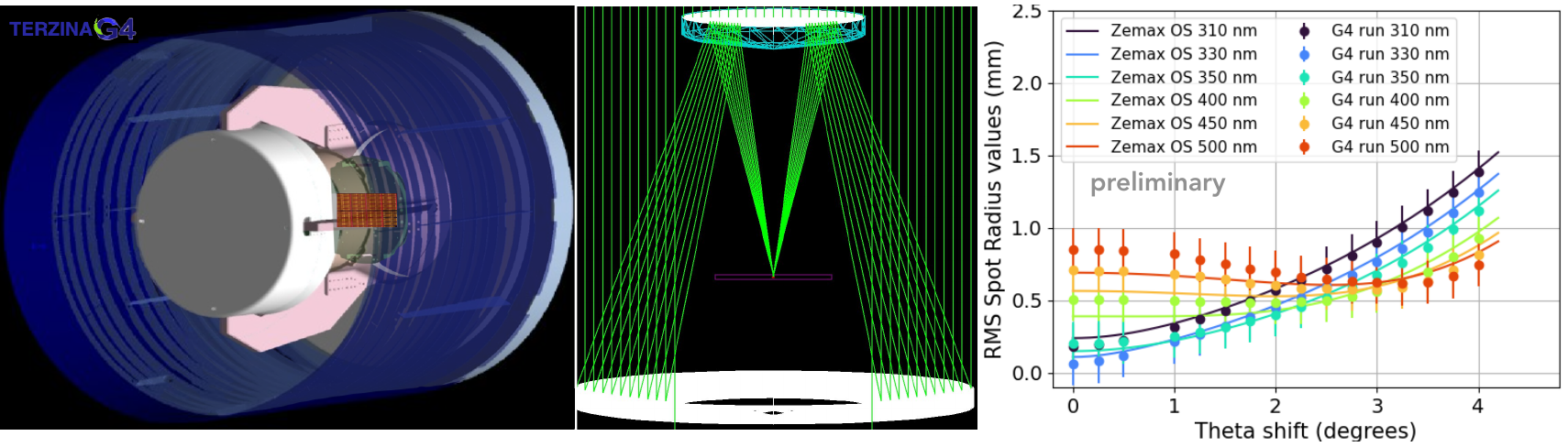}
    \caption{Implementation of a Schmidt–Cassegrain optical system in Geant4 (left). Comparison of the optical simulation results showing on-axis, off-axis spot sizes (right).}
    \label{fig:CombinedG4}
\end{figure}

The geometry of the Terzina telescope is simulated using Geant4. This step represents the first stage of the detector response, where the optical efficiency, shadowing effects from mechanical structures and pixel dead spaces are evaluated. The simulation can be run with different options depending on the study, including photon signal, background photons, radiation dose, and optical efficiency and Point Spread Function (PSF) evaluation. For photon simulations, the input consists of the photon flux density per m$^2$, including angular, temporal, and energy information. For radiation dose simulations, particle fluxes representative of the LEO environment are used, including direction and energy. In the photon case, the number of hits on the sensitive detector is stored, while in the radiation case, the energy deposited in relevant volumes is recorded. To achieve accurate and realistic simulations, both the mechanical structure and the optical system of the telescope have been realisticly modelled. The mechanical structure is imported from the CAD as tessellated solids and implemented in Geant4 via a custom \texttt{G4TessellatedSolid} module, accurately representing segmented and complex components (see the left panel in Figure~\ref{fig:CombinedG4}). In particular, the Terzina telescope consists of an Optical Head Unit that integrates three subsystems: the Optical Telescope Assembly (OTA), the Focal Plane Assembly (FPA), and a passive Thermal Control Assembly (TCA). The mechanical volumes of the OTA and FPA—including the optical bench, central support tower, baffles and vanes, and the FPA mechanical support—have been meshed, imported into the simulation, and assigned the corresponding material densities. The FPA houses the sensor plane and has been simulated in detail, with exact implementation of the 10 SiPM tiles arranged in two rows of five, including all dead spaces. Each array contains 8~$\times$~8 pixels, for a total of 640 pixels,  with individual pixel sizes of 3~$\times$~3~mm$^2$ and an active area near 2.73~$\times$ 2.34~mm$^2$ per pixel. Each pixel is treated as a sensitive volume where hits are collected. The mapping information of each pixel is used in the next step of the simulation. The final implementation of the real sensor plane is shown in Figure~\ref{fig:ReadoutSimFigures} (left side). The OTA features a hyperbolic primary mirror measuring 444~mm in diameter with a radius of curvature (RoC) of 1207~mm, along with an aspherical secondary mirror of 201.56~mm diameter and 626~mm RoC. To correct optical distortions (especially those caused by the flat focal plane) a 28 mm thick corrective lens with a 353~mm RoC is positioned before the secondary mirror. The detailed modelling of this optical system including the estimate of the PSF has led to the development of a dedicated optical library, \textsc{OpticsLibSim}\footnote{https://github.com/cattrima/OpticsLibSim.git}. In Figure~\ref{fig:CombinedG4}, the implementation of the optical system with a simulated on-axis beam (central panel) is shown, while on the right, the spot size of the optics, for different wavelengths and on-axis and off-axis, calculated with TerzinaG4 is compared to the Zemax results provided by the Officina Stellare company that designed the telescope’s optical system. It is remarkable to see how \textsc{OpticsSimLib}, specifically developed to model such complex optical systems in Geant4, is able to perfectly match the results obtained with the Zemax optical software.

\section{TerzinaReadoutSim}
\label{terzinareadoutsim}

\begin{figure}[ht]
    \centering

        \includegraphics[width=\textwidth]{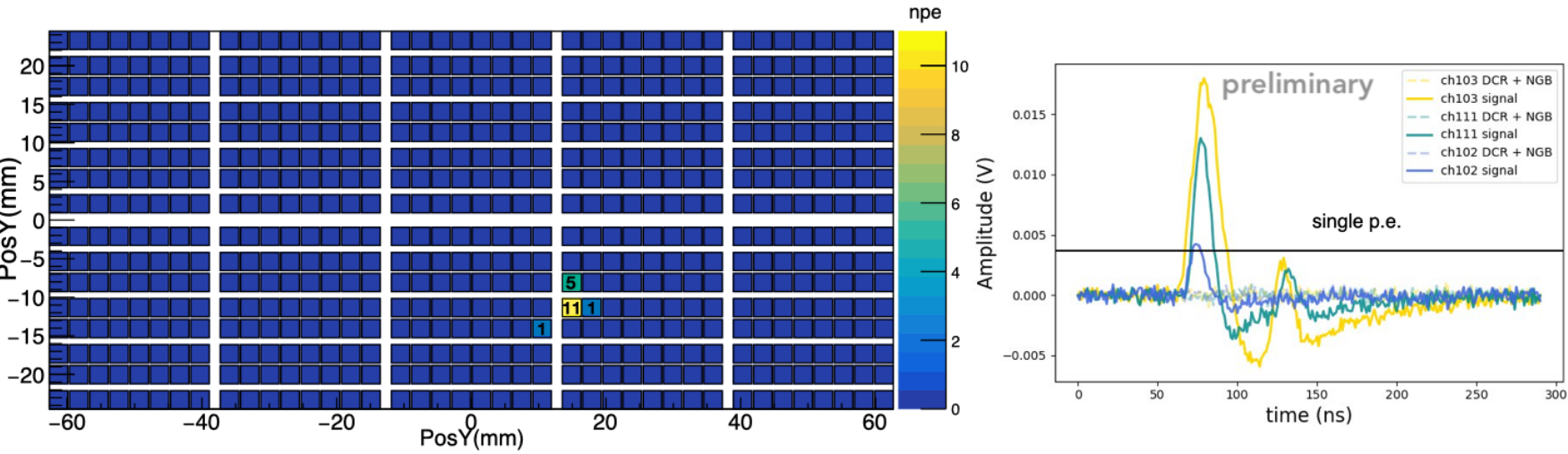}
 
    \caption{Simulation of the FPA response: focal plane hit pattern and corresponding waveforms of a simulated proton event at 300 PeV.}
    \label{fig:ReadoutSimFigures}
\end{figure}
The \textsc{TerzinaReadoutSim} module, fully integrated into the simulation framework, models the detector readout based on \textsc{TerzinaG4} photon hits and their channel-by-channel mapping on the camera plane. It incorporates the detector’s photon detection efficiency (PDE) and trigger response, and models the complete readout chain, including SiPM response with noise and gain variations, waveform generation, and front-end electronics, reproducing the time-dependent response to both signal and background sources. Taken together, these effects determine the final effective aperture of the detector.
The module requires several input parameters to operate correctly. These include the pulse template of the SiPM response to a single photoelectron (p.e.), scaled according to the over-voltage, as well as the probability of direct optical cross-talk (OCT) and after-pulse (AP) as a function of SiPM over-voltage, along with the after-pulse decay time. The root mean square error (RMSE) of the SiPM gain variation and the RMSE of electronic noise are also needed. The module uses the Night Glow Background (NGB) rate, derived from dedicated background studies, to configure trigger thresholds and set realistic noise levels, ensuring realistic signal-to-noise ratio (SNR). Similarly, the evolution of the dark count rate (DCR) as a function of temperature and cumulative radiation dose, obtained from Geant4 + SPENVIS simulations and SiPM characterisation \cite{JCAP_p}, is provided as input. Both NGB and DCR are included in the configuration and simulation code, enabling a fully realistic modelling of the detector response under operational conditions.
The SiPM response at microcell level is modelled by allowing each primary photoelectron to generate secondary photoelectrons through optical cross-talk (OCT) or after-pulsing (AP). The corresponding probabilities depend on the avalanche amplitude and history, and are reduced if the cell is not fully recharged (see \cite{leonid} for details).
Waveform simulations account for cosmic-ray induced air showers, background photons, single-photoelectron amplitudes, and statistical fluctuations. The DAQ system is modeled with a CITIROC-based trigger scheme. The trigger logic, implemented in both the ASICs and a dedicated FPGA, operates through two mechanisms: the Time Trigger and the Charge Trigger. 
Threshold parameters, coincidence, and time windows can be configured via an XML file; however, fixed thresholds are insufficient due to the dynamic observational background, including photon contributions from the Moon and city lights as well as radiation-induced variations in the SiPM dark count rate (DCR). An adaptive threshold approach is therefore applied, dynamically adjusting the trigger based on environmental conditions along the satellite orbit. The resulting photon and particle backgrounds, including temporal, angular, and radiation-induced effects, are incorporated into the readout and trigger simulation, providing realistic conditions for threshold settings. In Figure~\ref{fig:ReadoutSimFigures} the number of photoelectrons on the camera plane, together with the corresponding waveforms, is shown for a simulated proton shower at 300 PeV. The scientific results derived from this full simulation chain, including the final event selection, are presented in \cite{TerzinaICRC2025}.
\setlength{\parskip}{-10pt}
\section*{Acknowledgments}
NUSES is funded by the Italian Government (CIPE n. 20/2019), by the Italian Ministry of Economic Development (MISE reg. CC n. 769/2020), 
by the Italian Space Agency (CDA ASI n. 15/2022),  by the European Union NextGenerationEU under the MUR National Innovation Ecosystem  
grant ECS00000041 - VITALITY - CUP D13C21000430001 and by the Swiss National Foundation (SNF grant n. 178918). This study was carried out also in collaboration with the Ministry of University and  Research, MUR, under contract n. 2024-5-E.0 - CUP n. I53D24000060005. This research, leading to the  beam test results, also received partial funding from the European Union’s Horizon Europe research and innovation programme under grant agreement No. 101057511.


\begin{center}
{
\large 
\bf
The NUSES Collaboration 
}
\vspace{5ex}

M.~Abdullahi$^{a,b}$, R.~Aloisio$^{a,b}$, F.~Arneodo$^{c,d}$, S.~Ashurov$^{a,b}$, U.~Atalay$^{a,b}$, F.~C.~T.~Barbato$^{a,b}$, R.~Battiston$^{e,f}$, M.~Bertaina$^{g,h}$, E.~Bissaldi$^{i,j}$, D.~Boncioli$^{k,b}$, L.~Burmistrov$^{l}$, F.~Cadoux$^{l}$, I.~Cagnoli$^{a,b}$, E.~Casilli$^{a,b}$, D.~Cortis$^{b}$, A.~Cummings$^{m}$, M.~D'Arco$^{l}$, S.~Davarpanah$^{l}$, I.~De~Mitri$^{a,b}$, G.~De~Robertis$^{i}$, A.~Di~Giovanni$^{a,b}$, A.~Di~Salvo$^{h}$, L.~Di~Venere$^{i}$, J.~Eser$^{n}$, Y.~Favre$^{l}$, S.~Fogliacco$^{a,b}$, G.~Fontanella$^{a,b}$, P.~Fusco$^{i,j}$, S.~Garbolino$^{h}$, F.~Gargano$^{i}$, M.~Giliberti$^{i,j}$, F.~Guarino$^{o,p}$, M.~Heller$^{l}$, T.~Ibrayev$^{c,d,q}$, R.~Iuppa$^{e,f}$, A.~Knyazev$^{c,d}$, J.~F.~Krizmanic$^{r}$, D.~Kyratzis$^{a,b}$, F.~Licciulli$^{i}$, A.~Liguori$^{i,j}$, F.~Loparco$^{i,j}$, L.~Lorusso$^{i,j}$, M.~Mariotti$^{s,t}$, M.~N.~Mazziotta$^{i}$, M.~Mese$^{o,p}$, M.~Mignone$^{g,h}$, T.~Montaruli$^{l}$, R.~Nicolaidis$^{e,f}$, F.~Nozzoli$^{e,f}$, A.~Olinto$^{u}$, D.~Orlandi$^{b}$, G.~Osteria$^{o}$, P.~A.~Palmieri$^{g,h}$, B.~Panico$^{o,p}$, G.~Panzarini$^{i,j}$, D.~Pattanaik$^{a,b}$, L.~Perrone$^{v,w}$, H.~Pessoa~Lima$^{a,b}$, R.~Pillera$^{i,j}$, R.~Rando$^{s,t}$, A.~Rivetti$^{h}$, V.~Rizi$^{k,b}$, A.~Roy$^{a,b}$, F.~Salamida$^{k,b}$, R.~Sarkar$^{a,b}$, P.~Savina$^{a,b}$, V.~Scherini$^{v,w}$, V.~Scotti$^{o,p}$, D.~Serini$^{i}$, D.~Shledewitz$^{e,f}$, I.~Siddique$^{a,b}$, L.~Silveri$^{c,d}$, A.~Smirnov$^{a,b}$, R.~A.~Torres~Saavedra$^{a,b}$, C.~Trimarelli$^{a,b}$, P.~Zuccon$^{e,f}$, S.~C.~Zugravel$^{h}$.

 \vspace{5ex}

\begin{tabular}{c}
$^{a}$ Gran Sasso Science Institute (GSSI);\\ 
$^{b}$ Istituto Nazionale di Fisica Nucleare (INFN) - Laboratori Nazionali del Gran Sasso;\\ 
$^{c}$ Center for Astrophysics and Space Science (CASS);\\ 
$^{d}$ New York University Abu Dhabi, UAE;\\ 
$^{e}$ Dipartimento di Fisica - Università di Trento;\\ 
$^{f}$ Istituto Nazionale di Fisica Nucleare (INFN) - Sezione di Trento;\\ 
$^{g}$ Dipartimento di Fisica - Università di Torino;\\ 
$^{h}$ Istituto Nazionale di Fisica Nucleare (INFN) - Sezione di Torino;\\ 
$^{i}$ Istituto Nazionale di Fisica Nucleare (INFN) - Sezione di Bari;\\ 
$^{j}$ Dipartimento di Fisica M. Merlin dell’ Università e del Politecnico di Bari;\\ 
$^{k}$ Dipartimento di Scienze Fisiche e Chimiche -Università degli Studi di L’Aquila;\\ 
$^{l}$ Départment de Physique Nuclèaire et Corpuscolaire - Université de Genève, Faculté de Science;\\ 
$^{m}$ Department of Physics and Astronomy and Astrophysics, Institute for Gravitation and the Cosmos;\\ 
$^{n}$ Department of Astronomy and Astrophysics, University of Columbia;\\ 
$^{o}$ Istituto Nazionale di Fisica Nucleare (INFN) - Sezione di Napoli;\\ 
$^{p}$ Dipartimento di Fisica E. Pancini - Università di Napoli Federico II;\\ 
$^{q}$ now at The School of Physics, The University of Sydney;\\ 
$^{r}$ CRESST/NASA Goddard Space Flight Center;\\ 
$^{s}$ Dipartimento di Fisica e Astronomia - Università di Padova;\\ 
$^{t}$ Istituto Nazionale di Fisica Nucleare (INFN) - Sezione di Padova;\\ 
$^{u}$ Columbia University, Columbia Astrophysics Laboratory;\\ 
$^{v}$ Dipartimento di Matematica e Fisica “E. De Giorgi” - Università del Salento;\\ 
$^{w}$ Istituto Nazionale di Fisica Nucleare (INFN) - Sezione di Lecce. \\
\end{tabular}

\end{center}
\end{document}